\documentclass{article}

\usepackage{graphics}

\newcommand{\prl}[3]{{Phys. Rev. Lett.}{\bf #1}{(#2)}{#3}}
\newcommand{\prd}[3]{{Phys. Rev.}{\bf D#1}{(#2)}{#3}}
\newcommand{\pr}[3]{{Phys. Rev.}{\bf #1}{(#2)}{#3}}
\newcommand{\hepph}[1]{hep-ph/#1}
\newcommand{\hepex}[1]{hep-ex/#1}
\newcommand{\arnps}[3]{{Ann. Rev. Nucl. Part. Sci.}{\bf #1}{(#2)}{#3}}
\newcommand{\plb}[3]{{Phys. Lett.}{\bf B#1}{(#2)}{#3}}
\newcommand{\npb}[3]{{Nucl. Phys.}{\bf B#1}{(#2)}{#3}}
\newcommand{\mpla}[3]{{Mod. Phys. Lett.}{\bf A#1}{(#2)}{#3}}

\newcommand{\gae}{\stackrel{>}{\sim}}

\begin{titlepage}

\title{Top Theories\footnote{Presented at the 8th International Symposium on
    Heavy Flavour Physics, July 26-29, 1999, University of Southampton,
    Southampton, England.}}

\author{Elizabeth H. Simmons\thanks{e-mail address: simmons@bu.edu}\\
Department of Physics, Boston University\\
590 Commonwealth Avenue, Boston, MA  02215}

\begin{document}

\maketitle
\bigskip
\begin{picture}(0,0)(0,0)
\put(295,250){BUHEP-99-18}
\put(295,235){hep-ph/9908488}
\end{picture}
\vspace{24pt}
 
\begin{abstract}As the most recently discovered and heaviest quark, the top
  presents us with theoretical challenges.  How are we to understand its
  properties within the larger effort to explain the origins of electroweak
  and flavor symmetry breaking ?  This talk discusses some of the surprises
  the top quark may have in store for us and indicates how experiment may
  help us pinpoint the truth about top.
\pagestyle{empty}
\end{abstract}
\end{titlepage}

Since discovering the top quark in 1995 \cite{ehs-disc-top}, the CDF and D\O\ 
experiments have measured several of its properties with increasing
accuracy\footnote{For details about the status and prospects of top quark
  experiments, see the talk by G. Watts in these proceedings
  \cite{ehs-watts}}.  These include the top quark's mass, production
cross-section, and decay fraction to $b$ quarks.  The collaborations have
also begun studying characteristics such as the single-top production rate,
the kinematic distributions and spin correlations of pair-produced top
quarks, the helicity of $W$ bosons arising from in top decays, and the rate
of rare or non-standard top decays.

What impact will this information will have on particle theory?  Some
measurements will help us understand the top quark itself; others will help
us complete our understanding of the 3-generation standard model; still
others will be most informative about physics lying outside the standard
model.  This talk begins by discussing recent experimental studies of the top
quark in the context of the standard model.  Next, we review why it is
necessary to consider physics beyond the standard model.  The bulk of the
talk focuses on new physics that could be associated with the top quark and
how such physics might manifest itself in experiment.

\section{Top in the Standard Model}
\setcounter{equation}{0} 

Let us consider how two relatively well-measured properties of the top
quark, the mass and decay fraction to $b$ quarks, inform our understanding of
the standard model.

\subsection{Top Mass\protect\footnote{This discussion draws heavily upon \cite{ehs-willen}.}}

The mass of the top quark is measured to be 174 $\pm$ 5.1 GeV \cite{ehs-pdg}
. The large central value for the mass makes $m_t$ of order the weak scale.
This implies that the top quark's Yukawa coupling is of order 1, i.e. the
only Yukawa coupling of natural size.  The small size of the errors means
that the top quark has already tied the $b$ quark for the distinction of
best-measured fermion mass.  This is quite impressive, given that
measurements on the $b$ quark had nearly a 20-year head start!

\begin{figure}[ht]
{\scalebox{.35}{\includegraphics[-2.5in,1in][6in,8in]{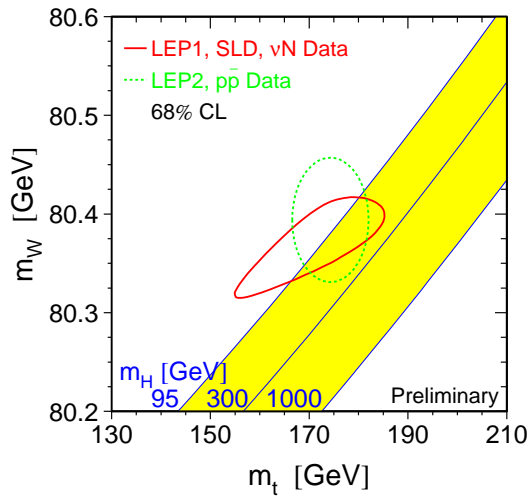}}\caption{Constraining
    $M_H$ using measured $m_t$ and $M_W$ \protect\cite{ehs-lepewwg}.  The
    closed curves show experimental limits on $m_t$ and $M_W$.  The values
    allowed in the standard model as a function of $M_H$ are indicated by the
    shaded band.}\label{fig1}}
\end{figure}

The precise measurement of the top quark mass makes it possible to combine
data on the top quark and $W$ boson masses in order to test the standard
model and constrain the mass of the Higgs boson.  The top quark, $W$ boson
and Higgs boson all contribute to radiative corrections to many observables
which have been well-measured at LEP or SLC or in low-energy neutrino
scattering.  Hence, given the measured values of the observables and the
experimental values of $M_W$ and $m_t$, it is possible to predict the range
within which the Higgs mass should lie.  As shown in figure \ref{fig1}, the
experimental constraints (closed curves) on $M_W$ and $m_t$ are consistent
with values of the Higgs boson mass allowed by the standard model (shaded
band).

In Run II at the Tevatron, it is anticipated that $m_t$ will be measured to
an accuracy of $\pm 3$ GeV (1 GeV in Run IIb) while $M_W$ should be measured
to 40 MeV by each experiment.  This level of precision will yield a
prediction of $M_H$ with $\delta M_H / M_H \leq 40\%$.  If the Higgs has not
still not been directly observed, this information will tell experiment where
to look!

\subsection{Decay Fraction to $b$ Quarks$^2$}

The top quark's decay fraction to $b$ quarks has been measured by CDF to be
\cite{ehs-bbmeas} 
\begin{displaymath}
B_b \equiv {\Gamma(t \to b W) \over {\Gamma(t \to q W)}} = 0.99 \pm 0.29 \, .
\end{displaymath}
Let us explore the significance of this number.

Within the three-generation standard model, $B_b$ is related to CKM matrix elements as follows
\begin{equation}
B_b \equiv {\vert V_{tb}\vert^2 \over {\vert V_{tb}\vert^2 + \vert V_{ts}\vert^2 + \vert V_{td}\vert^2}} \, .
\label{bbref}
\end{equation}
Three-generation unitarity implies that the denominator of (\ref{bbref}) is
precisely 1.0 .  Hence the measurement of $B_b$ tells us that \cite{ehs-bbmeas}
\begin{displaymath}
\vert V_{tb} \vert > 0.76\ \ \ \ \ (95\% c.l.)\ \, .
\end{displaymath}
However, within the 3-generation standard model, data on the light quarks
combined with CKM unitarity has already provided \cite{ehs-pdg} the much
tighter constraints $ 0.9991 < \vert V_{tb} \vert < 0.9994$, so that the
measurement of $\vert V_{tb} \vert$, while explicit, is not very informative.

If we extend the standard model by adding a fourth
generation of quarks, the analysis is rather different.  A search by D\O\ has
constrained \cite{ehs-pdg} the 4-th generation $b^\prime$ quark to have a
mass greater than $m_t - m_W$, so that the top quark could not readily decay
to $b^\prime$.  This means that the original expression for $B_b$
(\ref{bbref}) is still valid.  However, once there are four generations, the
denominator of the RHS of (\ref{bbref}) need not equal 1.0.  All we learn
from the CDF measurement of $B_b$ is that
\begin{displaymath}
\vert V_{tb} \vert \gg \vert V_{td} \vert\ , \ \vert V_{ts} \vert \, .
\end{displaymath}
On the other hand, light-quark data combined with 4-generation CKM unitarity
allows $\vert V_{tb}\vert$ to lie in the wide range $0.05 < \vert V_{tb}
\vert < 0.9994$ \cite{ehs-pdg}.  While the measurement of $B_b$ gives only
qualitative information about $\vert V_{tb}\vert$, that information is
nonetheless new and useful in the context of a 4-generation model.

Finally, we note that direct measurement of $\vert V_{tb} \vert$ in single
top-quark production at the Tevatron should reach an accuracy of 10\% in Run
IIa (5\% in Run IIb).  This will be quite useful in constraining physics
beyond the standard model.
 
\bigskip These two case studies show that each measured property of the top
quark may be expected to have multiple implications.  Their interpretation
depends strongly on the context provided by the underlying model of particle
physics.  Moreover, some data on the top quark may prove particularly
informative about physics {\it beyond} the standard model.

\section{Beyond the Standard Model}
\setcounter{equation}{0}

Two central concerns of particle theory are finding the cause of electroweak
symmetry breaking, which provides mass to the $W$ and $Z$ bosons, and
identifying the origin of flavor symmetry breaking, by which the quarks and
leptons obtain their diverse masses.  The standard model of particle physics,
based on the gauge group $SU(3)_c \times SU(2)_W \times U(1)_Y$ accommodates
both symmetry breakings by including a fundamental weak doublet of scalar
Higgs bosons ${\phi = {\phi^+ \choose \phi^0}}$ with potential function
$V(\phi) = \lambda \left({\phi^\dagger \phi - \frac12 v^2}\right)^2$.
However the standard model does not explan the dynamics
responsible for the generation of mass.

Furthermore, the scalar sector suffers from
\begin{figure}[ht]
{\scalebox{.36}{\includegraphics[-5.5in,0in][3in,1.25in]{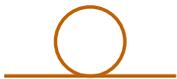}}
\caption{$ M_H^2\ \propto\ \Lambda^2$}\label{fig2}}
\end{figure}
\noindent 
two serious problems.  The scalar mass is unnaturally sensitive to the
presence of physics at any higher scale $\Lambda$ (e.g. the Planck scale), as
shown in figure \ref{fig2}.  This is known as the gauge hierarchy problem.
In addition, if the scalar must provide a good description of physics up to
arbitrarily high scale (i.e., be fundamental), the 
\begin{figure}[ht]
{\scalebox{.4}{\includegraphics[-5in,0in][3in,1.25in]{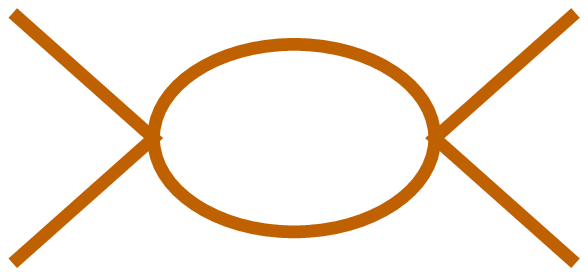}}
{\caption{$\beta(\lambda)\ = \ {{3\lambda^2}\over{2\pi^2}}\ > \
  0$}}\label{fig3}}
 \end{figure} 
\noindent scalar's self-coupling
($\lambda$) is driven to zero at finite energy scales as indicated in figure
\ref{fig3}.  That is, the scalar field theory is free (or ``trivial''). Then
the scalar cannot fill its intended role: if $\lambda = 0$, the electroweak
symmetry is not spontaneously broken.  The scalars involved in electroweak
symmetry breaking must therefore be composite at some finite energy scale.
We must seek the origin of mass in physics that lies beyond the standard
model and its fundamental scalar doublet.

One interesting possibility (denoted ``dynamical electroweak symmetry
breaking''\cite{ehs-1-dynam}) is that the compositeness of the scalar states
involved in electroweak symmetry breaking could manifest itself at scales not
much above the electroweak scale $v \sim 250$ GeV.  In these theories, a new
strong gauge interaction with $\beta < 0$ (e.g technicolor) breaks the chiral
symmetries of a set of massless fermions $f$ at a scale $\Lambda \sim 1$ TeV.
If the fermions carry appropriate electroweak quantum numbers, the resulting
condensate $\langle \bar f_L f_R \rangle \neq 0$ breaks the electroweak
symmetry as desired.  The logarithmic running of the strong gauge coupling
renders the low value of the electroweak scale (i.e.  the gauge hierarchy)
natural.  The absence of fundamental scalar bosons obviates concerns about
triviality.

\begin{figure}[hb] 

  {\scalebox{.5}{\includegraphics[-2in,0in][3in,1.5in]{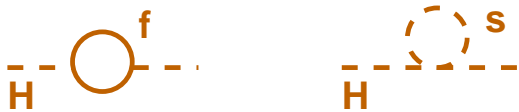}}\caption{$\delta
      M_H^2 \sim {g_f^2\over{4\pi^2}} (m_f^2 - m_s^2) + m_s^2 log
      \Lambda^2$}\label{fig4}}
\end{figure}

Another intriguing idea is to modify the standard model by introducing
supersymmetry \cite{ehs-1-susy}.  The gauge structure of the minimal
supersymmetric version of the standard model (MSSM) is identical to that of
the standard model, but each ordinary fermion (boson) is paired with a new
boson (fermion) called its ``superpartner,'' and two Higgs doublets are needed
to provide mass to the ordinary fermions.  As sketched in figure
\ref{fig4}, each loop of ordinary
particles contributing to the higgs boson's mass is now countered by a loop 
of superpartners.  If the masses of the ordinary particles and superpartners
are close enough, the gauge 
\noindent hierarchy can be stabilized \cite{ehs-1-stab}.
In addition, supersymmetry relates the scalar self-coupling to gauge
couplings, so that triviality is not a concern.

Once we are open to the idea of physics outside the standard model, the
question is where to seek experimental evidence.  Since the sample of top
quarks available for study in Run I at the Tevatron was relatively small, the
top quark may yet prove to have properties that set it apart from the other
quarks.  Examples include: light related states, low-scale compositeness, and
unusual gauge couplings.  The fact that $m_t$ is of order the weak scale
suggests that the top could even play a unique role in electroweak dynamics.
Upcoming top quark studies at the Tevatron's Run II will help us evaluate
these ideas.  for instance, a list of ``symptoms of new physics'' to look for
in the Run II top-pair sample is in \cite{ehs-demina}.

\section{Light Related States}
\setcounter{equation}{0} 

In many theories beyond the standard model, the spectrum of particles
accessible to upcoming experiments includes new states related to the top
quark. Some couple to the top quark, allowing the possibility of new
production or decay modes.  Others mix with the top quark, altering the
properties of the lighter ``top'' eigenstate we have seen relative to
standard model predictions.

\subsection{Light Top Squarks}

\begin{figure}[ht]
{\scalebox{.55}{\includegraphics[0in,1.75in][6in,7in]{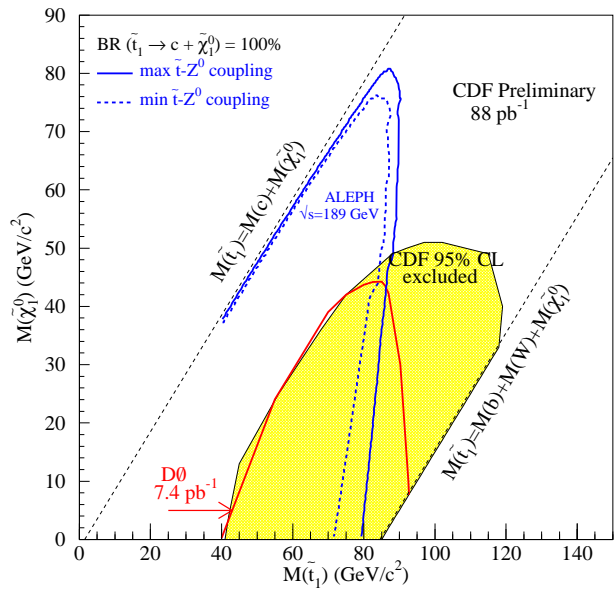}}
\caption{Searches for scalar top \protect\cite{ehs-3-slite} have excluded
  regions below the curves as shown, but still allow the stop to be lighter
  than the top .}\label{fig5}}
\end{figure}

Since supersymmetric models include a bosonic partner for each standard model
fermion, there is a pair of scalar top squarks affiliated with 
top  (one associated with $t_L$ and one, with $t_R$).  A glance at the
mass-squared matrix for the supersymmetric partners of the top quark:
\begin{center}
\begin{math}
\pmatrix{\tilde{M}^2_Q + m_t^2 &\ & 
{m_t}(A_t + \mu\cot\beta)\cr 
+ (M_Z^2\cos2\beta)\times &\ &\cr
(\frac12- \frac23\sin^2\theta_W) &\ &\cr
\ &\ & \tilde{M}^2_U + m_t^2 + \cr  
{m_t}(A_t + \mu\cot\beta)& &\frac23 M_Z^2 \sin^2\theta_W
    \cos2\beta\cr} 
\end{math}
\end{center}
reveals that the off-diagonal entries are proportional to $m_t$.  Hence, a
large top quark mass can drive one of the top squark mass eigenstates to be
relatively light.  Experiment still allows this possibility
\cite{ehs-3-slite}, as may be seen in figure \ref{fig5}.

Then perhaps some of the ``top'' sample observed in Run I included top
squarks \cite{ehs-3-stops}.  If the top squark is not much heavier than the top quark, it is
possible that $\tilde{t}\tilde{t}$ production occurred in Run I, with the top
squarks subsequently decaying to top plus neutralino ($\tilde{N}$) or gluino
$(\tilde{g})$.  On the other hand, if the top is a bit heavier than the stop,
some top quarks produced in $t\bar{t}$ pairs in Run I may have decayed to top
squarks via $t \to \tilde{t} \tilde{N}$ with the top squarks' subsequent
decay being either semi-leptonic $\tilde{t} \to b \ell \tilde\nu$ or
flavor-changing $\tilde t \to c \tilde{N}, c \tilde{g}$.  With either
ordering of mass, it is possible that gluino pair production occurred,
followed by $\tilde{g} \to t \tilde{t}$.

Such ideas can be tested by
studying the absolute cross-section, leptonic decays, and kinematic
distributions of the top quark events \cite{ehs-demina}.  For example, stop
or gluino production could increase the apparent $t\bar t$ production rate
above that of the standard model.  Or final states including like-sign
dileptons could result from gluino decays.

\subsection{Exotic quarks}

A variety of models propose the existence of a new charge 2/3 quark which
mixes with the top quark and alters the properties of the ``top'' state we
see from those predicted in the standard model.  In some models, the result
of the mixing is two nearly-degenerate states, which would imply that the top
sample at Run I contained an admixture of exotic quarks.
The larger top sample in Run II could make this apparent.  In other models,
the mass matrix of the top and its exotic partner is of a seesaw form
\begin{center}
\begin{math}
\pmatrix{\bar{t_L} & \bar{t}_L\prime\cr}
\pmatrix{0 & m_1 \cr m_2 & M \cr}
\pmatrix{t_R \cr t_R\prime \cr}
\end{math}
\end{center}
so that the extra state can be considerably heavier than the observed top
quark \cite{ehs-mix-lite}.  In this case, the best clue to the presence of
new physics might be alterations in the branching fractions of top quark
decays.

\subsection{Charged scalar bosons}

\begin{figure}[ht]
{\scalebox{.45}{\includegraphics[-1.75in,1.5in][6.5in,8.5in]
    {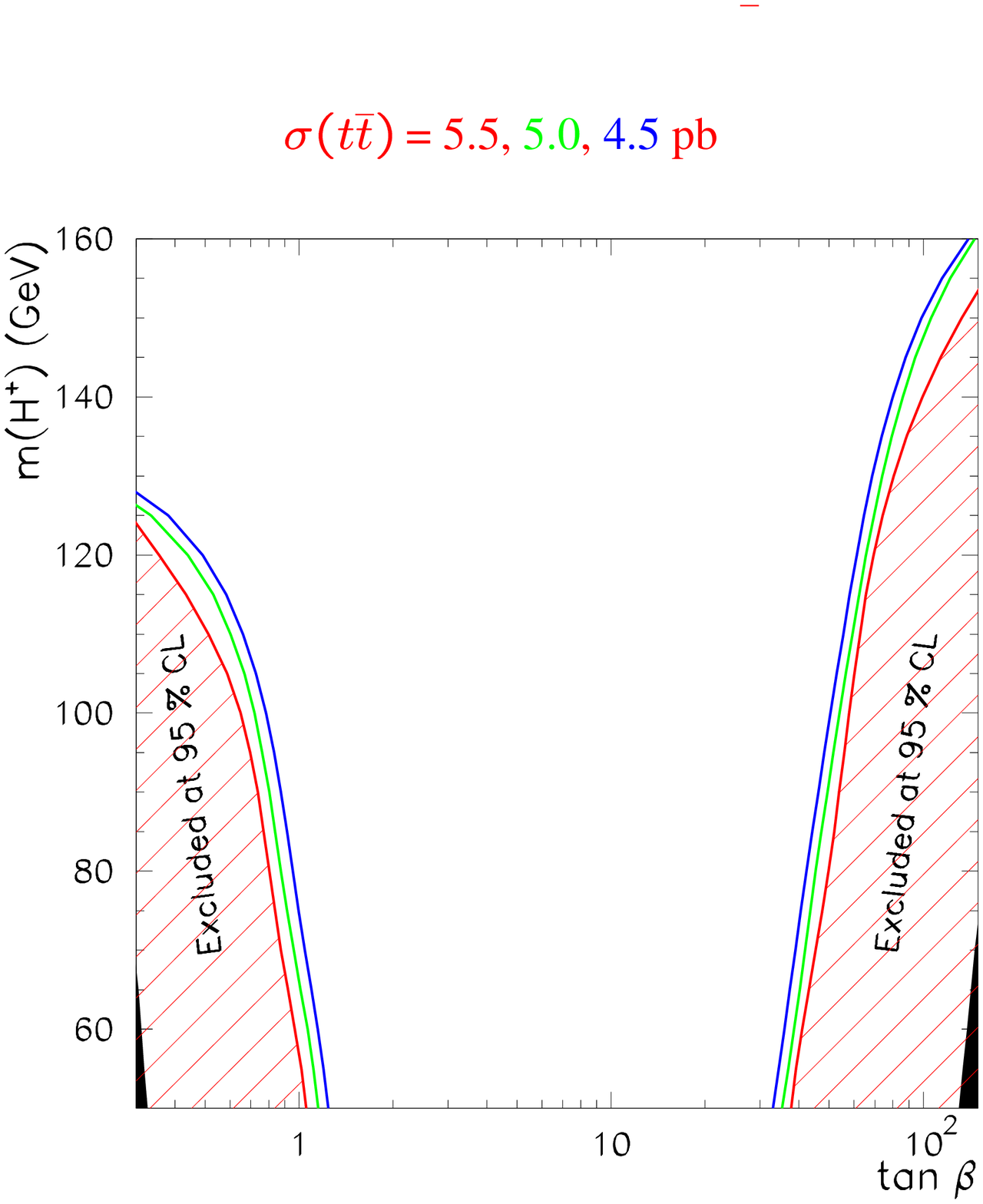}}\caption{D\O\ search for charged higgs bosons in top
    decays \protect\cite{ehs-higgs-plus}. The hatched regions of scalar mass
    and $\tan\beta$ are excluded. }\label{fig6}}
\end{figure}

Many quite different kinds of models include relatively light charged scalar
bosons, into which top may decay: $t \to \phi^+ b$.  SUSY models must include
at least two Higgs doublets in order to provide mass to both the up and down
quarks, and therefore have a charged scalar in the low-energy spectrum.  The
general class of models that includes multiple Higgs bosons likewise often
includes charged scalars that could be light.  Dynamical symmetry
breaking models with more than the minimal two flavors of new fermions (e.g.
technicolor with more than one weak doublet of technifermions) typically
possess pseudoGoldstone boson states, some of which can couple to third
generation fermions.  Run I data already limits the properties of light
charged scalars coupled to t-b (see figure \ref{fig6}); Run II will explore the
remaining parameter space still further.  

\subsection{FCNC decays}

If the large mass of the top arises from flavor non-universal
couplings between the top quark and new boson states, then
flavor-changing neutral current decays ($t \to c + X, u + X$) may result.
The Run I limits reported by CDF \cite{ehs-fcnc}
\begin{eqnarray}
BR(t \to Z q) &<& 0.33 \nonumber \\
BR(t \to \gamma q) &<& 0.032 \nonumber
\end{eqnarray}
leave ample room for new physics.

\section{Low-scale top compositeness}
\setcounter{equation}{0} 

\begin{figure}[ht]
{\scalebox{.25}{\includegraphics[-4in,0in][6in,4in]{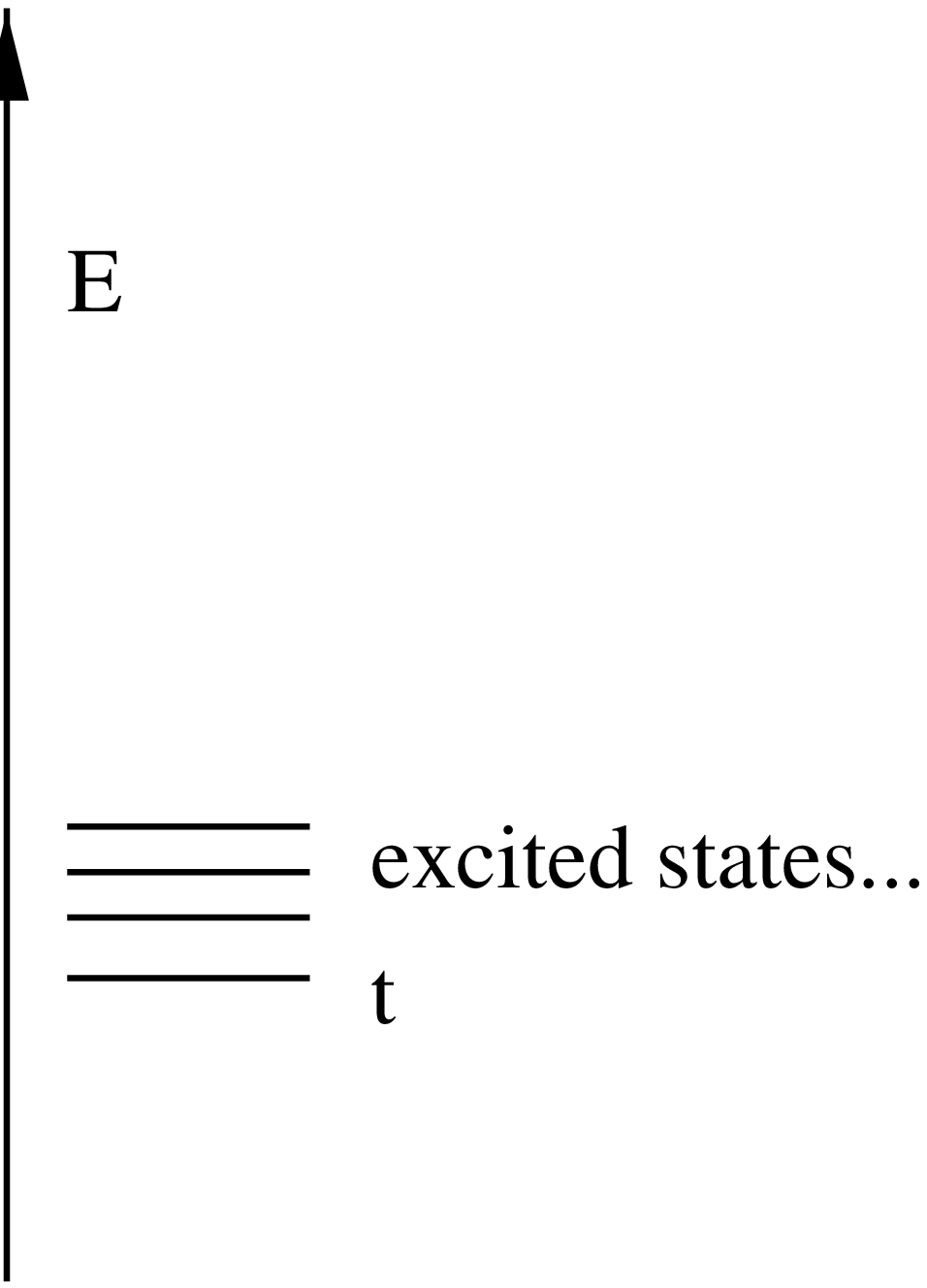}\hspace{1cm}
\includegraphics{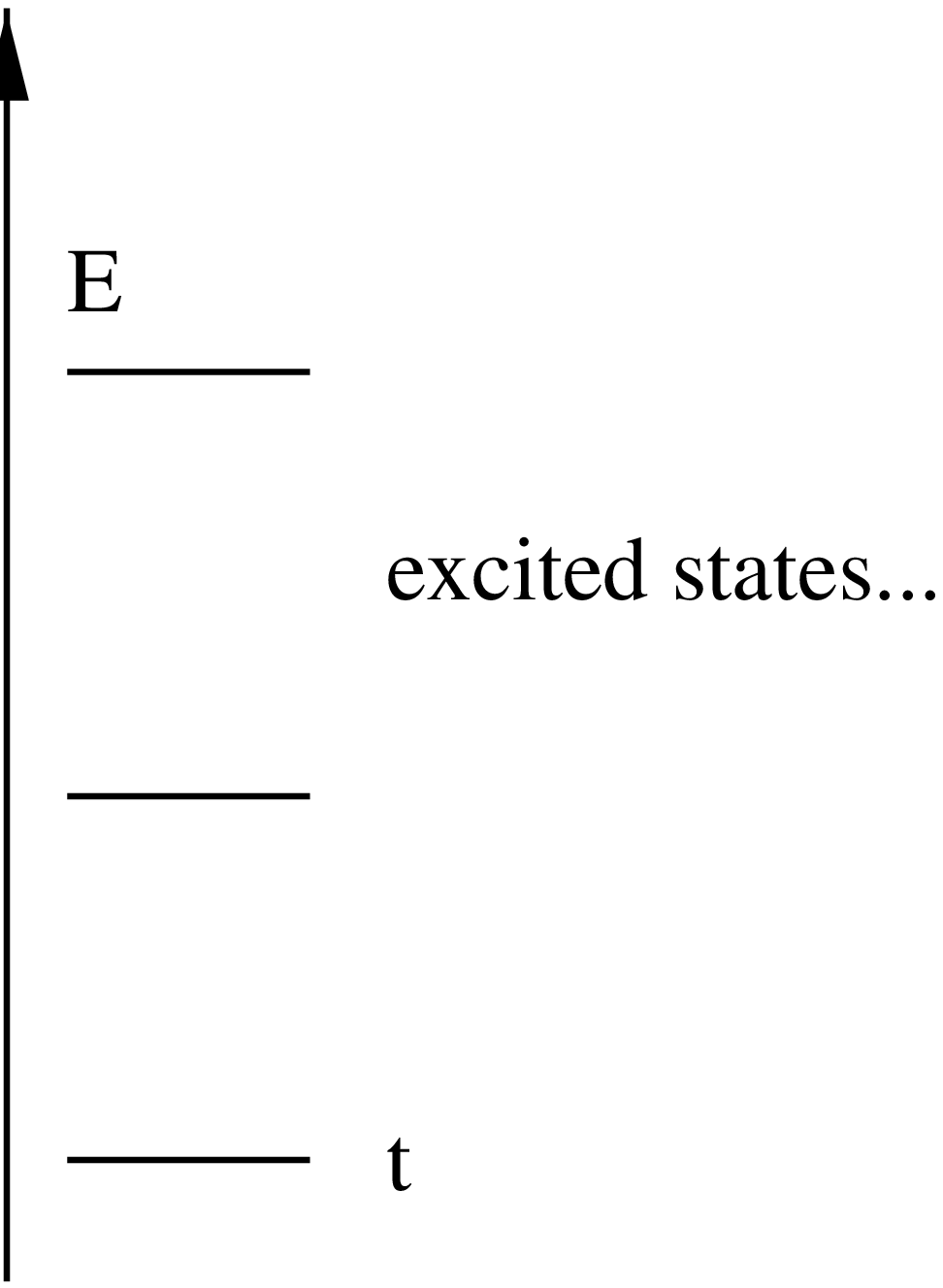}}\caption{A composite top quark would exhibit
  excited states.  Left:
      weak interactions underlying top compositeness produce inter-state
      spacing $\ll m_t$.  Right: strong interactions yield spacing $\gae
      m_t$.}\label{fig7}}
\end{figure}

We now turn to the possibility of a composite top quark.  Compositeness
requires new interactions to bind the consitutents together.  If those
interactions were weak, excited states of top would lie just above $m_t$;
strong coupling would produce large inter-state spacing (see figure
\ref{fig7}).  Since the three generations of quarks mix with one another, the
new interactions would couple at some level to first and second generation
quarks as well.  Thus, the absence of new weakly-coupled interactions of the
light fermions implies that top quark compositeness would have to arise from
strong interactions with a high intrinsic scale, $\Lambda$.

\begin{figure}[ht]{\scalebox{.25}{\includegraphics[-4.5in,0in][4in,3.5in]{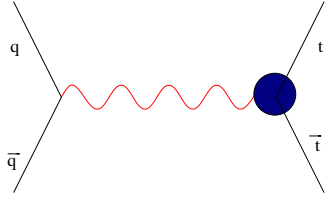}}
    \caption{  Composite top with colored constituents. $q\bar{q} \to t \bar{t}$
  scattering proceeds through gluon exchange:  $\sigma \approx \sigma_{SM} 
\left[ 1 + {\cal{O}}\left({\hat{s}\over\Lambda^2}\right)
\right]$}\label{fig8}}
\end{figure}

The magnitude of the effects of top compositeness on $q\bar{q} \to t \bar{t}$
depends on the properties of the constituents of the top.  If they carry color, scattering proceeds via gluon exchange and the 
cross-section is
modified from the QCD prediction by a form factor as in figure \ref{fig8}. This 
\noindent  possibility and related effects like anomalous top chromomagnetic
moments have been studied in \cite{ehs-top-chrom}.  If the light quarks are
also composite and share
\begin{figure}[ht]
{\scalebox{.25}{\includegraphics[-4.5in,0in][4in,4in]{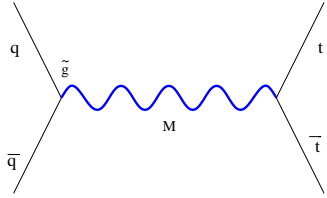}}
\caption{Composite  
      top and light quarks share constituents. $q\bar{q} \to t \bar{t}$
      scattering proceeds through interactions underlying compositeness: $\sigma
      \approx \sigma_{SM} \left[ 1 +
        {\cal{O}}\left({{\tilde{\alpha}\hat{s}}\over{\alpha_s M^2}}\right)
      \right] $}\label{fig9}}
\end{figure}
\noindent constituents with the top, scattering can
be caused directly by the interactions underlying compositeness (figure
\ref{fig9}) as well as by QCD gluon exchange.  As a result, the leading new
contributions to the scattering cross-section are enhanced by the strong
compositeness coupling, as envisaged in \cite{ehs-elp}.

\begin{figure}[hb]{\scalebox{.3}{\includegraphics[-4in,0in][4in,5.5in]{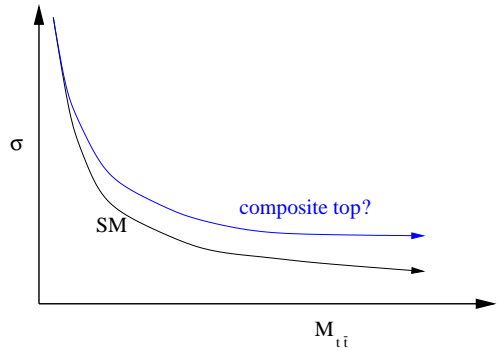}}\caption{Schematic
      invariant mass distribution of pair-produced top quarks in the standard
      model (SM) and assuming composite top quarks.}} \end{figure}

In either case, the form of the effect on the invariant mass distribution is
the same: increased events at high invariant mass. The experimental reach (in
$\Lambda$ or $M/\tilde{g}$) of the Run II experiments remains an open
question.  For example, it is not clear whether a deviation large enough to
be seen would be consistent with the cross-section's respecting unitarity.
Also worth further study is the possibility that a helicity analysis of the
produced $t \bar{t}$ pairs could reveal the form of the interactions
underlying top compositeness.

\section{Unusual quantum numbers}
 \setcounter{equation}{0}
 
 A top quark participating in physics beyond the standard model could have
 new gauge quantum numbers.  Recent model-building has proposed several
 extensions of the standard model gauge groups that treat third-generation
 quarks (and sometimes leptons) differently from the lighter fermions.  In
 this section, we illustrate the possibilities and note how one might test
 them experimentally.  Section 6 shows a more complete model built using
 these principles.

\subsection{Extended Strong Interactions}
\setcounter{equation}{0}

One interesting possibility is to extend the strong interactions in a way
that causes them to distinguish among fermion flavors at energies 
above the weak scale.  At high energies, the strong interactions would then
include both an $SU(3)_H$ for the $t$ (and $b$) and an $SU(3)_L$ for the
other quarks.  to be consistent with low-energy hadronic data, these groups
must spontaneously break to their diagonal subgroup
(identified with $SU(3)_{QCD}$) at a scale $M$:
\begin{displaymath}
SU(3)_H \times SU(3)_L \to SU(3)_{QCD}\, .
\end{displaymath}
As a result of the symmetry breaking, a color octet of heavy gauge bosons
preferentially coupled to $t$ and $b$ is present in the spectrum at scales
below $M$.  

The extra gauge bosons have useful theoretical consequences.  Exchange of
\begin{figure}[hb]
{\scalebox{.45}{\includegraphics[0in,1.9in][6in,7.5in]{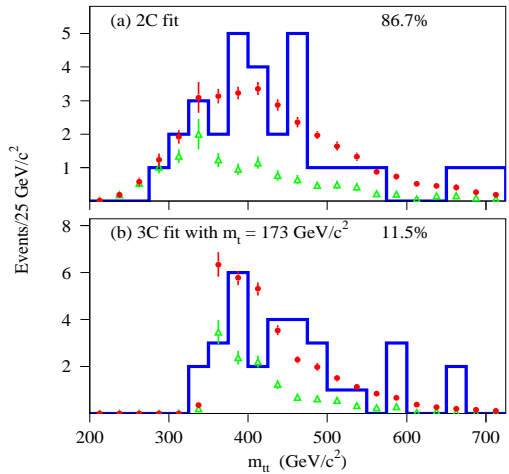}}
  \caption{Top-pair invariant mass spectrum from D\O\
    \protect\cite{ehs-watts}.  The 
    histogram 
    shows the data.  The open triangles are Monte Carlo background; the solid
    dots are MC background plus top signal.}\label{figd}} 
\end{figure}
the heavy gauge bosons yields a new four-fermion interaction
\begin{displaymath}
- {4\pi\kappa \over M^2}
\left(\overline{t}\gamma_\mu {\lambda^a\over 2} t\right)^2
\end{displaymath}
that can cause top quark condensation ($ \langle \bar{t}t\rangle \neq 0$)
\cite{ehs-3-ttcond} .
This provides an opportunity for dynamical symmetry breaking to provide a
large mass for the top quark.  Furthermore, because the new interaction
treats top and bottom quarks identically, it need not make an unacceptably large
contribution to $\Delta\rho$.

\begin{figure}[h]
{\scalebox{.45}{\includegraphics[0in,3.5in][6in,9in]{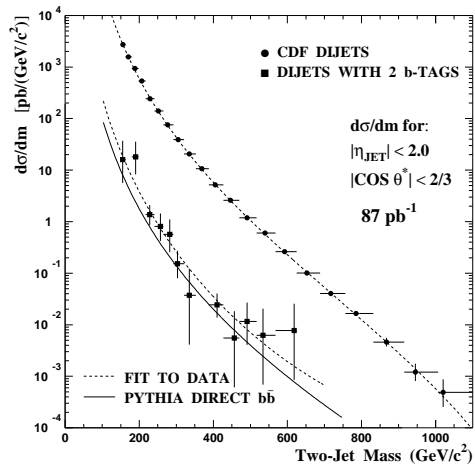}}
\caption{CDF dijet and $b\bar{b}$ spectra compared with PYTHIA standard model
    predictions \protect\cite{ehs-cdf-jets}.}\label{figc}} \end{figure}

Experimental tests of the extended strong interactions can be based on the
fact that the extra colored gauge bosons that become massive in this model
couple preferentially to the top and bottom quarks.  One may therefore, as
CDF \cite{ehs-cdf-jets} and D\O\ are already doing, seek evidence of new
resonances in the 
$t \bar{t}$ or $b\bar{b}$  invariant mass
spectrum (figures \ref{figd} and \ref{figc}) that do not also appear in 
the (light) dijet invariant mass spectrum.

\subsection{Extended Hypercharge Interactions}

A second possibility is to extend the hypercharge group to include a $U(1)_H$
felt by third-generation fermions and a $U(1)_
L$ felt by the light fermions.  Again, this extended group must be broken
at some high energy scale to its diagonal subgroup, which is identified with
the standard $U(1)_Y$:
\begin{displaymath}
U(1)_H \times U(1)_L \to U(1)_Y  \, .
\end{displaymath}
In the context of new strong dynamics, an extended hypercharge interaction
can be used to help generate the observed large splitting between the masses
of the top and bottom quarks, because these quarks carry
different values of hypercharge (see Section 6).

\begin{figure}[ht]
{\scalebox{.35}{\includegraphics[-2in,2in][5in,6in]{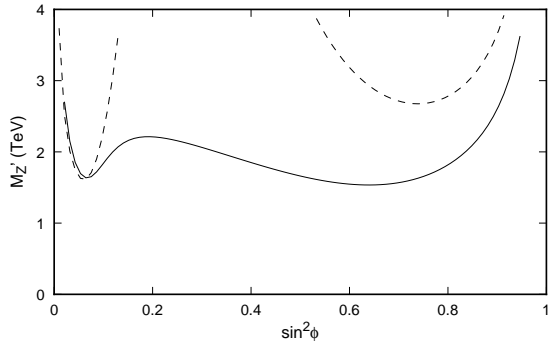}}
\caption{Lower bound on $Z'$ mass as a function of $Z$ $Z'$ 
  mixing angle $\phi$ \protect\cite{ehs-3-rscjtbd}.}\label{fig14}} 
\end{figure}

\begin{figure}[ht]
{\scalebox{.45}{\includegraphics[-2in,2.5in][6in,9in]{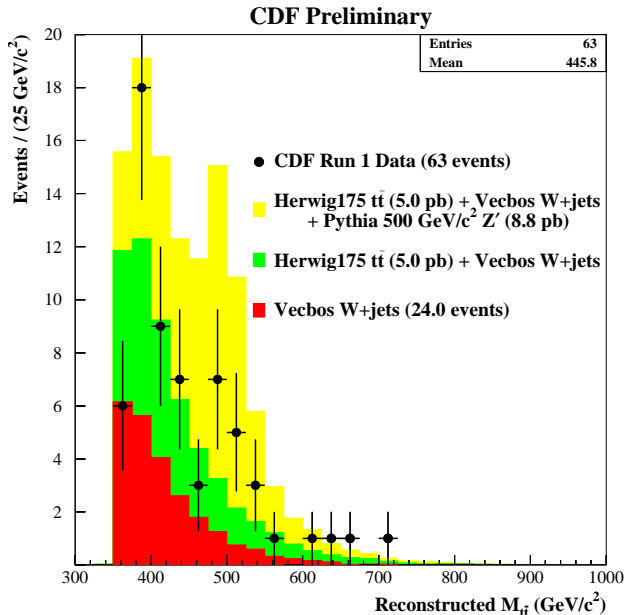}}\caption{CDF
    search for $Z'$ in the $t\bar{t}$ spectrum. Dots are data. Histograms are
    Monte Carlo: red is background, green includes top signal, yellow
    indicates effects of including a $Z'$ boson
    \protect\cite{ehs-watts}.}\label{fig14a}}\end{figure}

The broken hypercharge generator manifests itself physically as a heavy $Z'$
boson.  Indirect searches for such a $Z'$ look in precision low-energy and
$Z$-pole data for evidence of its mixing
with the ordinary $Z$.  A lower bound of 1.5 - 2 TeV on the
mass of the $Z'$ \cite{ehs-3-rscjtbd} has been set in this way (see figure
\ref{fig14}).  Direct
searches for a $Z'$ boson that couples preferentially to the third-generation
fermions can also be made in the invariant mass spectra of $t\bar{t}$,
$b\bar{b}$ and $\tau^+ \tau^-$.  Preliminary searches in the $t\bar{t}$
spectrum by CDF show no signs of a $Z'$ (see figure \ref{fig14a}); the search
will continue at Run II.

\subsection{Extended Weak Interactions}
Alternatively, one might extend the weak gauge group to include an $SU(2)_H$
felt by third-generation fermions and an $SU(2)_L$ coupled to the light
fermions \cite{ehs-2-ncetc}\cite{ehs-topfl}.  To preserve approximate weak
universality at low energies, this extended group must be spontaneously
broken at a high energy scale to its diagonal subgroup, which is identified
with the standard $SU(2)_W$:
\begin{displaymath}
SU(2)_H \times SU(2)_L \to SU(2)_W  \, .
\end{displaymath}
Because the breaking of the weak gauge group is central to generating
fermion masses, separation of the weak interactions of the heavy and light
fermions can allow distinct origins for their masses.  This can
help circumvent some of the traditional difficulties with constructing
dynamical models of mass generation.  

A class of dynamical models of this type \cite{ehs-2-ncetc}, called
``non-commuting extended technicolor'' (NC-ETC), has the symmetry-breaking
pattern
\begin{eqnarray*}
G_{ETC}  &\times& SU(2)_{L}\\
&\downarrow& \\
G_{TC} \times SU(&\!\!\!2&\!\!\!)_{H}  \times SU(2)_{L} \\
&\downarrow&\\
G_{TC}  &\times& SU(2)_{W}
\end{eqnarray*}
in which $SU(2)_H$ is embedded in the ETC interactions at high energies.
Cancellation between the effects of ETC gauge boson exchange and mixing
between the $Z$ bosons of the two $SU(2)$ groups enables $R_b$ to have a
value consistent with experiment.  At the same time, weak boson mixing causes
the weak interactions of the top quark to differ from those of the up and
charm quarks at low energies.

\begin{figure}[ht]
{\scalebox{.45}{\includegraphics[-1.5in,3.5in][7.5in,7in]
    {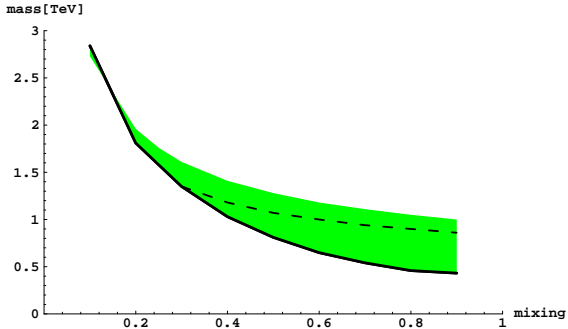}}\caption{The vertical axis is the heavy $W$'s mass; the
    horizontal axis is the degree of mixing of the two weak gauge groups in a
    non-commuting ETC model.  The area below the solid curve is excluded by
    precision electroweak data \protect\cite{ehs-2-ncetc}.  In the shaded
    region, $R_\sigma$ would be increased by at least 16\%
    \protect\cite{ehs-3-wtb2}. }\label{fig15}}
\end{figure}

Non-standard top quark weak interactions may be detectable in single
top-quark production at Run IIb \cite{ehs-3-wtb}\cite{ehs-3-wtb2}.  The ratio
of cross-sections $R_\sigma \equiv \sigma(\bar{p}p \to t b) / \sigma(\bar{p}p
\to l \nu)$ can be measured (and calculated) to an accuracy
\cite{ehs-3-wtbacc} of at least $\pm 8\%$.  In NC-ETC models, mixing of the
$W$ bosons from the two weak groups alters the light $W$'s coupling to the
final-state fermions, including top quarks.  So long as the heavy $W$ bosons
are not too massive, the resut is a visible {\bf increase} in $R_\sigma$ (see
Figure \ref{fig15}).

\section{Unique Role in Electroweak Dynamics}
\setcounter{equation}{0} 

New physics associated with the top quark will
be most interesting if it helps explain electroweak symmetry breaking.  If
top squarks are discovered in the Run II ``top'' sample, one reason for
enthusiasm would be a first sighting of particles outside the standard model
spectrum; but even more important would be the proof that low-energy
supersymmetry must be included in any non-standard physics that seeks to
explain the origin of mass. If the reaction $t \to \phi^+ b$ is
observed in Run II, the immediate question will be ``Is $\phi^+$ a
Higgs or a technipion ?''.

In some theories, the top quark itself helps explain the origin of mass.
Those in which the top quark has new gauge interactions are of particular
interest, because they can help resolve some outstanding difficulties of the
original dynamical electroweak symmetry breaking scenarios.  A key challenge
for models of dynamical mass generation is to provide simultaneously
\newline \null \hspace{.5cm} $\bullet$ the correct $M_W$ and $M_Z$, with
$\Delta\rho \approx 0$ \cite{ehs-2-dirr}
\newline \null \hspace{.5cm} $\bullet$ both $m_t$ and $m_t - m_b$ large
\newline \null \hspace{.5cm} $\bullet$ $R_b$ near the standard model
value. \cite{ehs-2-sbs} 
\newline The original extended technicolor models have difficulty meeting
this challenge.  Dynamical models with extended weak interactions
\cite{ehs-2-ncetc} have more success, but no complete model has been
constructed.  Here, we focus on dynamical models with extended strong (and,
sometimes, hypercharge) interactions, known as ``topcolor-assisted
technicolor'', which have made progress on all three issues.


The prototypical topcolor-assisted technicolor model \cite{ehs-3-tc2}
has the following gauge group and symmetry-breaking pattern.
\begin{eqnarray*}
G_{TC} &\times& SU(2)_W \times \\
{U(1)_H \times U(1)_L} &\times& {SU(3)_H \times SU(3)_L}  \\
&\downarrow& \ \ M \gae 1 {\rm TeV} \\
G_{TC} \times SU(2)_{W} &\times&
  {U(1)_Y} \times {SU(3)_C} \\
&\downarrow&\ \ \ \Lambda_{TC}\sim 1 {\rm TeV} \\
{U(1)_{EM}} &\times& {SU(3)_{C}}\, .
\end{eqnarray*}
The groups $G_{TC}$ and $SU(2)_W$ are
ordinary technicolor and weak interactions; the strong and hypercharge
groups labeled ``H'' couple to 3rd-generation fermions and have stronger
couplings than the ``L'' groups coupling to light fermions
The separate $U(1)$ groups ensure that the bottom quark will not
condense when the top quark does.  Below the scale $M$, the Lagrangian
includes effective interactions for $t$ and $b$:
\begin{eqnarray} 
&-&{{4\pi \kappa_{tc}}\over{M^2}} \left[\overline{\psi}\gamma_\mu  
{{\lambda^a}\over{2}} \psi \right]^2 \\
&-&{{4\pi \kappa_1}\over{M^2}} \left[{1\over3}\overline{\psi_L}\gamma_\mu  
\psi_L + {4\over3}\overline{t_R}\gamma_\mu  t_R
-{2\over3}\overline{b_R}\gamma_\mu  b_R
\right]^2 \, . \nonumber
\end{eqnarray}
So long as the following relationship is satisfied (where the critical
value is $\kappa_c \approx 3\pi/8$ in the NJL approximation \cite{ehs-3-njl}) 
\begin{displaymath}
\kappa^t = \kappa_{tc} +{1\over3}\kappa_1 >
\kappa_c  >
\kappa_{tc} -{1\over 6}\kappa_1 = \kappa^b\, ,
\end{displaymath}
only the top quark will condense and become very massive \cite{ehs-3-tc2}.

The topcolor-assisted technicolor models combine the strong points of
topcolor and extended technicolor scenarios to give a more complete dynamical
picture of the origin of mass
features\cite{ehs-3-tc2phen}\cite{ehs-1-dynam}. Technicolor 
causes most of the electroweak
symmetry breaking, with the top condensate contributing a decay constant
$f \sim 60$ GeV; this prevents $\Delta\rho$ from being too large, as
mentioned earlier.  So long as the $U(1)_H$ charges of the
technifermions are isospin-symmetric, they cause no additional large
contributions to $\Delta\rho$.  ETC dynamics at a scale $M \gg 1$TeV
generates the light fermion masses and contributes about a GeV to the
heavy fermions' masses; this does not generate large corrections to
$R_b$.  Finally, the top condensate provides the bulk of the top quark
mass and the top-bottom splitting.  The unique role of the top quark is what
makes these models of mass generation viable.

\section{Conclusions}

Top quark studies at the Tevatron's Run I have provided our first look at the
top quark and given some insight into its properties within the standard
model.  Run II will clearly enable us to learn far more.  The quest for
understanding electroweak symmetry breaking and fermion masses points to
physics beyond the Standard Model.  This opens the possibility that the top
quark may have unusual characteristics, some of which could become apparent
during Run II.  Whether the new physics associated with top is compositeness,
new related states, new gauge interactions, or something not yet imagined (!)
it would be tremendously exciting if it also helped reveal the origins of
mass.

\bigskip \centerline{\bf Acknowledgments} 

Thanks are due to S. Willenbrock for discussions on the top in the
standard model and to R.S. Chivukula for discussions on top compositeness.


\end{document}